\documentclass{kapproc} 

\let\footnote\savefootnote

\def\lesssim{\mathrel{\hbox{\rlap{\hbox{\lower4pt\hbox{$\sim$}}}\hbox{$<$}}}}
\def\gtrsim{\mathrel{\hbox{\rlap{\hbox{\lower4pt\hbox{$\sim$}}}\hbox{$>$}}}}

\usepackage{procps} 

\usepackage[dvips]{graphicx}

\upperandlowercase

\kluwerbib 

\begin{document}
\def\aj{AJ}%
\def\araa{ARA\&A}%
\def\apj{ApJ}%
\def\apjl{ApJ}%
\def\apjs{ApJS}%
\def\ao{Appl.~Opt.}%
\def\apss{Ap\&SS}%
\def\aap{A\&A}%
\def\aapr{A\&A~Rev.}%
\def\aaps{A\&AS}%
\def\azh{AZh}%
\def\baas{BAAS}%
\def\jrasc{JRASC}%
\def\memras{MmRAS}%
\def\mnras{MNRAS}%
\def\pra{Phys.~Rev.~A}%
\def\prb{Phys.~Rev.~B}%
\def\prc{Phys.~Rev.~C}%
\def\prd{Phys.~Rev.~D}%
\def\pre{Phys.~Rev.~E}%
\def\prl{Phys.~Rev.~Lett.}%
\def\pasp{PASP}%
\def\pasj{PASJ}%
\def\qjras{QJRAS}%
\def\skytel{S\&T}%
\def\solphys{Sol.~Phys.}%
\def\sovast{Soviet~Ast.}%
\def\ssr{Space~Sci.~Rev.}%
\def\zap{ZAp}%
\def\nat{Nature}%
\def\iaucirc{IAU~Circ.}%
\def\aplett{Astrophys.~Lett.}%
\def\apspr{Astrophys.~Space~Phys.~Res.}%
\def\bain{Bull.~Astron.~Inst.~Netherlands}%
\def\fcp{Fund.~Cosmic~Phys.}%
\def\gca{Geochim.~Cosmochim.~Acta}%
\def\grl{Geophys.~Res.~Lett.}%
\def\jcp{J.~Chem.~Phys.}%
\def\jgr{J.~Geophys.~Res.}%
\def\jqsrt{J.~Quant.~Spec.~Radiat.~Transf.}%
\def\memsai{Mem.~Soc.~Astron.~Italiana}%
\def\nphysa{Nucl.~Phys.~A}%
\def\physrep{Phys.~Rep.}%
\def\physscr{Phys.~Scr}%
\def\planss{Planet.~Space~Sci.}%
\def\procspie{Proc.~SPIE}%
\let\astap=\aap
\let\apjlett=\apjl
\let\apjsupp=\apjs
\let\applopt=\ao



\articletitle[]{The Birth of Massive Stars\\and Star Clusters}











--------------

\author{Jonathan C. Tan}
\affil{Institute of Astronomy, Dept. of Physics, ETH Z\"urich, 8093 Z\"urich, Switzerland}
\email{jt@phys.ethz.ch}








\begin{abstract}
  In the present-day universe, it appears that most, and perhaps all,
  massive stars are born in star clusters. It also appears that all
  star clusters contain stars drawn from an approximately universal
  initial mass function, so that almost all rich young star clusters
  contain massive stars. In this review I discuss the physical
  processes associated with both massive star formation and with star
  cluster formation. First I summarize the observed properties of
  star-forming gas clumps, then address the following questions. How
  do these clumps emerge from giant molecular clouds? In these
  clustered environments, how do individual stars form and gain mass?
  Can a forming star cluster be treated as an equilibrium system or is
  this process too rapid for equilibrium to be established? How does
  feedback affect the formation process?
\end{abstract}



\section{Introduction}

Star clusters\footnote{I define a star cluster as a group of stars
  that forms together from a gravitationally bound gas {\it clump}.},
are the fundamental units of star formation in galaxies. Most Galactic
stars are born in clusters (Lada \& Lada 2003): their figure~2 implies
that equal numbers of stars are forming in each logarithmic interval
of cluster mass, for cluster masses from $\sim 50 - 1000\:M_\odot$.
There is a dearth of star formation in clusters below $50\:M_\odot$.
The sample of Lada \& Lada (2003) is too small to constrain the
initial cluster mass function beyond $\sim 1000\:M_\odot$.  Hubble
Space Telescope observations have probed this range in external
galaxies, finding a continuation of the mass function slope (Larsen
2002). For the dwarf starburst galaxy NGC~5253, Tremonti et al. (2001)
have proposed a model in which all star formation occurs in clusters,
which then dissolve on timescales of $\sim 10$~Myr to create the
sources of the observed diffuse UV light.

The initial mass function of stars in clusters appears largely
invariant (Kroupa 2002) so that almost all relatively massive clusters
will contain at least a few high-mass stars. Thus a significant fraction
of all stars form in proximity to massive stars, and may be affected
by their strong feedback.

Locally, essentially all massive stars form in clusters (de Wit et al.
2005), so high-mass star formation seems to require an environment
that will also produce a large number and mass of low-mass stars. In
the present-day universe, massive star formation and star cluster
formation are one and the same process.

It is clear that an understanding of massive star and star
cluster formation is important to many areas of astrophysics, from
galaxy evolution to planet formation.

\section{Overview of physical properties}

Figure~1 shows the masses, $M$, and mean surface densities,
$\Sigma=M/(\pi R^2)$, of star clusters and interstellar gas clouds.
For convenience $\Sigma = 1\:{\rm g\:cm^{-2}}$ corresponds to
$4800\:M_\odot\:{\rm pc^{-2}}$, $N_{\rm H}=4.3\times 10^{23}\:{\rm
  cm^{-2}}$ and $A_V=200$~mag, for the local gas to dust ratio.
Contours of constant radial size, $R$, and hydrogen number density, $n_{\rm
  H} = \rho/\mu = 3M / (4\pi R^3 \mu)$, where $\mu=2.34\times
10^{-24}\:{\rm g}$ is the mean mass per H, are indicated. The density
contours also correspond to free-fall timescales, $t_{\rm
  ff}=\sqrt{3\pi/(32G\rho)}=1.38\times 10^{6} (n_{\rm H}/10^3\:{\rm cm^{-3}})^{-1/2}\:{\rm yr}$. For a virialized cloud with virial
parameter $\alpha_{\rm vir}\simeq 1$ (Bertoldi \& McKee 1992) the
signal crossing or dynamical timescale is $t_{\rm dyn}=2.0t_{\rm ff}$.

The presence of molecules allows interstellar gas to cool to low
temperatures, $\sim 10-20~\:{\rm K}$, effectively removing thermal
pressure support. To survive the destructive local
interstellar FUV radiation field requires a total column of $N_{\rm
  H}=(0.4,2.8)\times 10^{21}\:{\rm cm^{-2}}$ for $\rm H_2$ and CO,
respectively (McKee 1999). Giant molecular clouds (GMCs) have an
approximately constant column of $N_{\rm H}=(1.5\pm0.3)\times
10^{22}\:{\rm cm^{-2}}$ and typical masses $\sim 10^5-10^6\:M_\odot$
(Solomon et al. 1987). A sample of local ($d\lesssim 3$~kpc) infrared-dark
clouds (IRDCs), discussed in \S3, have masses ranging from several
hundred to $\sim 10^4\:M_\odot$ and $\Sigma\sim 0.1\:{\rm g\:cm^{-2}}$
(Kirkland \& Tan, in prep.), about a factor of 3 greater than the mean
value of GMCs. The massive star forming clumps observed in the sub-mm
by Mueller et al. (2002) have similar masses, but surface densities
typically a factor of five greater still. More revealed star clusters,
such as the Orion Nebula Cluster, have similar properties. More
massive and higher surface density clusters are rare, but can be found
in the Galactic center, e.g. the Arches and Quintuplet clusters (e.g.
Kim et al. 2000). The most massive young clusters, so-called super
star clusters, are often found in starburst environments,
such as the Antennae galaxies, and in some dwarf galaxies, e.g.
NGC~5253 (Turner et al. 2000) and NGC~1569 (Gilbert \& Graham 2003).

All high-mass star-forming systems appear to be at about a constant
density of $n_{\rm H} \sim 2\times 10^5\:{\rm cm^{-3}}$, corresponding
to $t_{\rm ff}\sim 1 \times 10^5\:{\rm yr}$. This is about the same as
the density at which the cooling rate is a maximum (Larson 2005), and
thus gravitational collapse is easiest. A spherical
self-gravitating cloud in hydrostatic equilibrium with mean surface
density $\Sigma$ and density profile $\rho\propto r^{-k_\rho}$ with
$k_\rho=1.5$, similar to observed clumps, has a mean pressure of
$4.3 \times 10^8\Sigma^2\:{\rm K~cm^{-3}}$ (McKee \& Tan
2003). Massive stars and star clusters appear to form under
pressures $\gtrsim 3\times 10^7\:{\rm K~cm^{-3}}$, much
higher than that of the local diffuse ISM, i.e. $2.8\times
10^4\:{\rm K~cm^{-3}}$ (Boulares \& Cox 1990).

\begin{figure}[h]
\includegraphics[width=\textwidth]{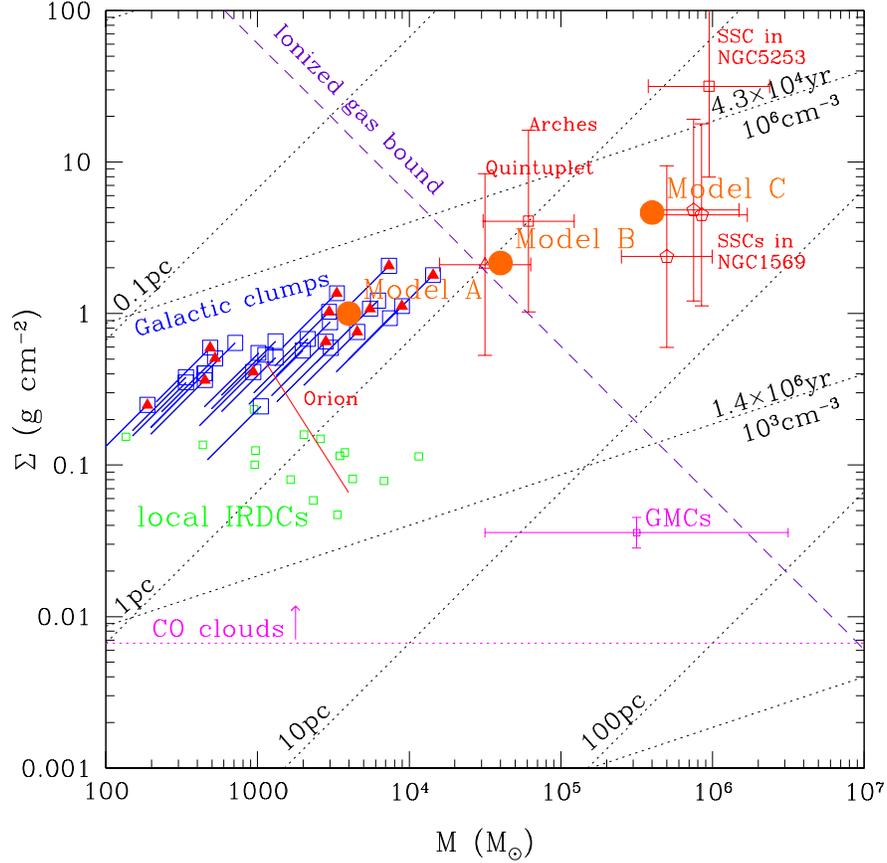}
\caption{Surface density, $\Sigma$, versus mass, $M$, for star clusters 
  and interstellar clouds. Contours of constant radius , $R$, and
  hydrogen number density, $n_{\rm H}$, or free-fall timescale, $t_{\rm
    ff}$, are shown with dotted lines. The minimum surface density for
  CO clouds in the local Galactic UV radiation field is shown, as are
  typical parameters of GMCs. Dense, cold clumps of GMCs known as
  infrared dark clouds are shown by small open squares (see text in
  \S3).  Large open squares are the star-forming clumps of Mueller et
  al.  (2002): a triangle indicates the clump contains an HII region
  while the diagonal line from each point shows the effect of
  uncertain dust opacities on the mass estimate. The Orion Nebula
  Cluster, allowing for a contribution from gas of 50\%, is shown by
  the solid diagonal line, which traces conditions from the inner to
  the outer parts of the cluster. Several more massive clusters are
  also indicated (see \S2 for references). The condition for ionized
  gas to remain bound is indicated by the dashed line. The three solid
  circles are the conditions of feedback models discussed in \S6.}
\end{figure}

\section{Setting up initial conditions for star cluster formation}

What causes a particular region of a GMC to form a star cluster? From
Figure~1 we see that the surface density, pressures, and volume
densities must increase by at least factors of 10, 100, and 1000,
respectively. This occurs in only a small part of the GMC: typically
only $\sim$1\% of the mass is involved.

Models for the cause of star formation can be divided into two groups:
quiescent and triggered. In the former, star formation occurs in the
densest, most unstable clumps of the GMC, and these form out of
the general gravitational contraction of the entire cloud. This
process may be regulated by the decay of turbulence, ambipolar
(Mouschovias 1996) or turbulent diffusion of magnetic flux, or
heating and ionization (McKee 1989) from newly-formed star clusters.
In models of triggered star formation, the star-forming clumps are
created by compression of parts of the GMC by external causes, such as:
cloud collisions (Scoville et al. 1986; Tan 2000); convergent
turbulent flows (e.g. Mac Low \& Klessen 2004); 
or feedback from young stars with ionization (Elmegreen \& Lada 1977;
Thompson et al. 2004; Deharveng et al. 2005), stellar winds (e.g.
Whitworth \& Francis 2002), protostellar outflows, radiation pressure,
and supernovae (e.g. Palous et al. 1994).

Elmegreen (2004) has noted that the compressions that result from most
forms of stellar feedback are probably only efficient within
particular GMCs or GMC complexes, i.e. young stars forming in one GMC
are unlikely to trigger star formation in another. Oey et al. (2005)
claim the age sequence of 3 regions of the W3/W4 complex is evidence
for triggered star formation over an approximately $100$~pc scale
region.


A promising method for addressing the cause of star cluster formation
is the study of infrared-dark clouds (IRDCs). These are regions that
have surface densities high enough to obscure the Galactic infrared
background. Large numbers of IRDCs have been found towards the inner
Galaxy with the Midcourse Space Experiment (MSX) (Egan et al. 1998).
These are associated with dense molecular gas (Carey et al.  1998;
Teyssier, Hennebelle \& P\'erault 2002). Carey et al. (1998) measured
the following physical properties: radii $r\sim 0.2-8$~pc, column
densities $\Sigma\sim 0.5 - 50\:{\rm g\:cm^{-2}}$, densities $n_{\rm
  H}\gtrsim 2\times 10^5\:{\rm cm^{-3}}$, masses $M\sim 10^3 -
10^5\:M_\odot$ and temperatures $T\lesssim 20$~K. Kirkland \& Tan (in
prep.) identified a sample of relatively nearby IRDCs from the MSX
infrared survey and the Galactic Ring Survey of $\rm ^{13}CO$
(Simon et al. 2001).  Surface densities and masses were estimated with
three independent methods: infrared extinction, line strengths of $\rm
^{13}CO$, and virial arguments.  The dispersion between these
methods is a factor of a few due to systematic errors. The average
properties of each cloud are shown in Figure~1. They have similar
masses to the star clusters and surface densities that are somewhat
smaller. Thus they are likely to be representative of the earliest
stage of star cluster formation.

From a visual inspection of the sample, Kirkland \& Tan (in prep.)
find that the morphologies of the IRDCs are more varied, and in
particular filamentary, than the star-forming clumps, that are often
approximately spherical (Shirley et al. 2003). The line widths are
several km/s, which is much greater than the sound speed of gas at
$\sim 20$~K. Thus the initial conditions for star cluster formation
probably involve supersonic turbulence, although not necessarily
super-Alfv\'enic turbulence. There are often multiple velocity
components. Some IRDCs are relatively isolated from
other star-forming regions suggesting that their formation
does not require triggering by feedback from young stars.

\section{How do stars form within clusters?}

We have seen that star clusters are born from turbulent gas, i.e.
having velocity dispersions much greater than thermal. A basic
question is how individual stars form in this environment. In
particular do they grow inside quasi-equilibrium gas {\it cores} that
collapse via accretion disks with relatively stable orientations? In
this scenario (e.g. Shu, Adams, \& Lizano 1987, McKee \& Tan 2003) the
initial mass of the core helps to determine the final mass of the
star, modulo the effects of protostellar feedback.  Alternative models
involving competitive Bondi-Hoyle accretion (e.g. Bonnell et al. 2001)
and direct stellar collisions (Bonnell, Bate, \& Zinnecker 1998) have
been proposed. These alternative models have been particularly
motivated for the case of massive star formation since this occurs in
the most crowded regions, radiation pressure feedback from massive
stars on dust grains can cause problems for standard accretion
scenarios, and the Jeans mass in these high pressure, high density
regions is only a fraction of a solar mass.

\subsection{Formation of Cores}

First consider the formation of cores from a turbulent medium.
Ballesteros-Paredes, Klessen, \& V\'azquez-Semadeni (2003) and Klessen
et al. (2005) find that a substantial fraction of ``cores'' identified
in their nonmagnetic SPH simulations of supersonic turbulence appear
to be quiescent (i.e. line widths $\leq$ than thermal) and coherent
(i.e. line widths are roughly independent of positional offset from
the core center), but are in fact dynamic, transient entities. They
argue that the inference of hydrostatic equilibrium, e.g. from radial
profiles that appear similar to Bonnor-Ebert profiles (e.g. Alves,
Lada, \& Lada 2001), is not necessarily valid, since such profiles are
also possible for dynamically evolving cores. However, it is not clear
if these artificial dynamic cores are consistent with the observations
of Walsh et al. (2004), which find very small ($\lesssim 0.1\:{\rm
  km\:s^{-1}}$) velocity differences between the line centers of high
($n_{\rm H} \sim 4\times 10^{5}\:{\rm cm^{-3}}$) and low ($n_{\rm H}
\sim 2\times 10^{3}\:{\rm cm^{-3}}$) density traces of starless cores:
real cores do not appear to be moving with respect to their envelopes.
Estimates of the ages of starless cores (e.g. Crapsi et al. 2005) are
uncertain, but have the potential to constrain models of core
formation.

The numerical simulations described above are nonmagnetic. Li et al.
(2004) and V\'azquez-Semadeni et al. (2005) have studied the
properties of cores forming from turbulent, magnetized gas. The latter
authors find in their periodic, fixed grid, isothermal, ideal MHD,
driven turbulence simulations, that: magnetic fields reduce the
probability of core formation; in the magnetically subcritical run, a
bound core forms that lasts $\sim 5 t_{\rm ff}$ (defined at densities
$\sim 50$ times the mean), which would be enough for ambipolar
diffusion to affect the dynamics; in the moderately supercritical
case, where magnetic fields are relatively weaker, bound cores form
and then are able to undergo runaway collapse over about $2t_{\rm
  ff}$, defined at the core's mean density. These results suggest that
the initial conditions for star formation are bound cores, and that
the stronger the magnetic field, the more chance the cores have to
attain a quasi-equilibrium structure. The marginally critical case is
probably most relevant if star-forming clumps evolve from regions of
GMCs that gradually lose magnetic support. The observations by
Crutcher (1999) of the magnetic field strength in dense regions of
GMCs imply that these regions are only marginally supercritical and
that magnetic fields are important for the dynamics.

Magnetic fields are likely to affect the masses of cores that are
present in a given environment. One argument against massive star
formation from cores has been that the thermal Jeans mass in the high
pressure, high density regions associated with massive star formation
is very small. However this argument is irrelevant if massive cores
derive most of their pressure support from either magnetic fields or
turbulent motions. Observations suggest that the mass function of
cores is fairly similar, within large uncertainties, to that of stars
and that there are some massive pre-stellar cores (Testi \& Sargent
1998; Motte et al. 2001; Li, Goldsmith, \& Menten 2003; Beuther \&
Schilke 2004).

\subsection{Accretion to Stars}

It is computationally expensive to follow gravitational collapse to
the high densities and short timescales associated with protostars and
their accretion disks. A common numerical technique is to introduce
sink particles in bound regions of high density, which can then
accrete gas from their surroundings (Bate, Bonnell, \& Price 1995).
Bonnell \& Bate (2002) modeled star cluster formation with SPH,
isothermal, non-periodic, no feedback, nonmagnetic simulations, with
initial setup of static gas and sink particles about to undergo global
collapse.  Stars gained mass via competitive accretion and stellar
collisions and the final mass spectrum was similar to the Salpeter
mass function.  Using similar simulations, except now with an
initially turbulent velocity field with no later driving, Bonnell,
Vine, \& Bate (2004) showed that the most massive star at the end of
their calculation had gained mass that was initially very widely
distributed. Dobbs, Bonnell, \& Clark (2005) found that a massive
turbulent core, such as envisaged by McKee \& Tan (2003), can fragment
into many smaller cores and protostars if the equation of state is
isothermal.  However, their non-isothermal model suffered much less
fragmentation, while the results of V\'azquez-Semadeni et al. (2005)
suggest that less fragmentation would also occur if magnetic fields
are allowed to affect the dynamics.  Schmeja \& Klessen (2004)
simulated star cluster formation with SPH simulations with periodic
boundaries, driven turbulence, no magnetic fields, no feedback, sink
particle diameters of 560~AU, and an isothermal equation of state,
finding highly variable accretion rates for their protostars.

We have seen that SPH simulations, by lacking magnetic fields,
probably do not accurately model the fragmentation process of real
star-forming regions, particularly with regard to core formation.
Another difficulty is that in SPH simulations with sink particles,
``stars'' acquire most of their mass by competitive Bondi-Hoyle
accretion and this process is not adequately resolved. In theory, gas
is gravitationally focused by a passing star so that streamlines
collide, shock and dissipate their energy. Eulerian grid simulations,
including sink particles (Krumholz, McKee, \& Klein 2004) and adaptive
mesh refinement of small scale structures, have been used to simulate
the interaction of sink particles with surrounding turbulent gas: the
accretion rate is much smaller than the classical analytic estimate of
accretion from a uniform medium (Krumholz, McKee, \& Klein 2005a).
Stellar feedback should also reduce this accretion rate, particularly
to massive stars (e.g. Edgar \& Clarke 2004). Thus the importance of
Bondi-Hoyle accretion may be grossly over-estimated in SPH
simulations.

\subsection{Assumptions and Predictions of the McKee-Tan Model}

McKee \& Tan (2002; 2003) modeled massive star formation by assuming
an initial condition that is a massive core in approximate pressure
equilibrium with the surrounding protocluster medium, i.e. the
star-forming clump. Tan \& McKee (2002) modeled star cluster formation
by extending this assumption to every star. To derive the pressure in
the clump, the system was assumed to be in approximate hydrostatic
equilibrium so that the mean pressure is related to the surface
density, i.e. $P\sim G\Sigma^2$.  How valid are these assumptions?

The mean pressure in the clump sets the overall density normalization
of each core and thus its collapse time and accretion rate. The
McKee-Tan model allows for deviations from exact pressure equilibrium
with the parameter $\phi_P$, although the expectation is that these
deviations will be factors of order unity.  Although the core is
treated as collapsing in isolation, this is also an approximation:
McKee \& Tan (2003) estimate that during the collapse time the core
will interact with an amount of mass similar to the initial core mass,
although not all of this will become bound to the core. Thus in
reality one would expect for a given initial core mass somewhat
greater final stellar masses than under the assumption of isolated
cores. The particular density structure of cores assumed by McKee \&
Tan (2003) is $\rho \propto r^{-k_\rho}$ with $k_\rho=1.5$ set from
observed cores. This choice affects the evolution of the accretion
rate during the collapse: $k_\rho<2$ implies accretion rates
accelerate. However, this is a secondary effect compared to the
overall normalization of the accretion rate that is set by the
external pressure. In any case since the pressure support is
nonthermal with significant contributions from turbulent motions, one
does not expect a smooth density distribution in the collapsing core,
and the accretion rate will show large variations about the mean.

The assumption of approximate pressure equilibrium in the protocluster
requires star formation to occur over at least several dynamical
timescales, and this is examined in the next section. The basic
picture of star cluster formation then involves: a turbulent,
self-gravitating gas clump in which bound cores occasionally form
(most gas at any given time is not in bound, unstable cores); a core
mass function fairly similar to stars, i.e. massive cores form but are
rare; an approximate equilibrium of cores with their surroundings; the
collapse of cores quite rapidly in one or two free-fall timescales to
form stars or binaries; the orbiting of newly-formed stars in the still
star-forming clump, but negligible growth via competitive accretion.

Some of the key predictions of the McKee-Tan model are the properties
of the cores and accretion disks of massive stars. The core size is
$R_{\rm core} \simeq 0.06 (M_{\rm
  core}/60M_\odot)^{1/2}\Sigma^{-1/2}\:{\rm pc}$.  Recall that
$\Sigma$, the surface density of the clump, is related to the pressure
of clump via $P\sim G\Sigma^2$. These small, pressure-confined cores
have relatively small cross-sections for close interactions with other
stars, although such interactions may still become important in the
later stages of cluster formation once the stellar density has been
built up to a high enough level. The rate of core collapse leading to
accretion to the star, via a disk, is $\dot{m}_* = 4.6\times 10^{-4}
f_*^{1/2} M_{60}^{3/4}\Sigma^{3/4}\:M_\odot\:{\rm yr}^{-1}$, where
$f_*$ is the ratio of $m_*$ to the final stellar mass and a 50\%
formation efficiency is assumed. Thus the collapse time, $1.3\times
10^5 M_{60}^{1/4} \Sigma^{-3/4}\:{\rm yr}$, is short and quite
insensitive to $M$, allowing coeval high and low mass star formation.
The disk size is $R_{\rm disk}=1200 (\beta/0.02) (f_* M_{60})^{1/2}
\Sigma^{-1/2}{\rm AU}$, where $\beta$ is the initial ratio of
rotational to gravitational energy of the core, and the normalization
is taken from typical low-mass cores (Goodman et al. 1993), although
there is quite a large dispersion about this value. These estimates
allow quantitative models of the protostellar evolution, disk
structure and outflow intensity.  These have been compared to
observations of the Orion KL protostar (Tan 2004a), also discussed
briefly below. First I review other observational evidence of massive
star formation from cores and accretion disks.

\subsection{Observational Evidence for Massive Star Formation from Cores and Accretion Disks}

The issue of the mode of star formation, particularly massive star
formation, is most likely to be resolved by observations. What
observations are required? A common approach has been to search for
disks around massive stars. However, these by themselves do not
distinguish between the models, unless they are seen in conjunction
with a collapsing pre-stellar core and it can be shown that the star
accumulated most of its mass by accretion from the core through the
disk. One would also like to show that the disk has maintained a
fairly stable orientation, perhaps by looking at the impact of past
outflow activity, during the accretion process, although even this is
not necessarily to be expected from the collapse of a very turbulent
core.

There are a number of claims for disks around massive protostars and
massive young stars.  Cesaroni et al. (1999) made mm and IR
observations of IRAS~20126+4104, concluding the system showed the
expected signatures of a massive ($\sim 24\:M_\odot$) protostar,
forming from a Keplerian accretion disk inside a dense gas core.
Shepherd, Claussen, \& Kurtz (2001) used 7~mm observations to
marginally resolve the driving source, G192.16, of a powerful
molecular outflow, which from luminosity arguments is thought to be a
$\sim 10\:M_\odot$ protostar. They interpreted the elongation, which
is roughly perpendicular to the outflow, to be evidence for a $\sim
100$~AU, $\sim 10\:M_\odot$ disk. However, much of the elongation is
asymmetric, and so they also invoked a second protostar.  Sandell,
Wright, \& Forster (2003) used 3.4~mm continuum and molecular line
observations of NGC7538S to infer the presence of a rotating, massive
($\sim 100\:M_\odot$), and exceptionally large ($r_d \sim 14000$~AU)
disk about a $\sim 10^4\:L_\odot$ protostar, again driving a powerful
outflow. This source is also peculiar in that, if it is a massive
protostar, it is relatively isolated. Beltr\'an et al.  (2004) used
1.4~mm continuum and molecular line observations to identify 4 massive
protostellar disks by searching for velocity gradients perpendicular
to outflows. They found disk sizes of several thousand AU. Chini et
al. (2004) used NIR imaging and CO line observations in M17 to find an
elongated structure $\sim2-3\times 10^4$~AU across with a mass of
$\gtrsim 100\:M_\odot$ and a velocity gradient of $1.7\:{\rm
  km\:s^{-1}}$. In the above systems the velocities measured from
molecular lines are typically on quite large scales that barely
resolve the disk: the velocity differences are only a few $\rm
km\:s^{-1}$, since the inner regions are not resolved. It is possible
that some of these sources, particularly those where there is little
evidence for a luminous central source or outflow, may simply be
flattened or filamentary structures with a velocity gradient.

Pestalozzi et al. (2004) interpreted VLBI observations of methanol
masers in NGC7538~IRS~N1 in terms of an edge-on Keplerian disk
extending to a radius of $\sim 1000$~AU and orbiting a $30\:M_\odot$
protostar. While some methanol maser systems may trace accretion
disks, it appears that many are in fact signatures of outflows (De
Buizer 2003).

In more evolved and revealed systems, NIR spectra of CO and Br$\gamma$
emission have been used to infer the presence of disks,
the emission coming from inside a few AU from the star (Blum et al.
2004; Bik \& Thi 2004). Vink et al. (2002) used H$\alpha$
spectropolarimetry to show that Herbig Ae/Be stars are surrounded by
flattened, presumably disk-like, structures. While studies of revealed
systems are useful for probing the properties and lifetimes of remnant
accretion disks, they do not directly test the different formation
scenarios, since even stellar collisions would be expected to leave
remnant material that would form a disk.

Most of the aforementioned systems are at distances of $\sim 2$~kpc or
more. The closest massive protostar is in the Orion KL region, only
$\sim 450$~pc away. Wright et al. (1996), Greenhill et al. (1998) and
Tan (2004) have interpreted the system as containing a $r\sim 1000$~AU
accretion disk, as traced by SiO (v=0; J=2-1) maser emission, centered
about the thermal radio source {\it I} (Menten \& Reid 1995) and
aligned perpendicular to the large scale molecular outflow that flows
to the NW and SE. However, from SiO (v=1,2; J=1-0) masers within
several tens of AU from source {\it I}, Greenhill et al. (2003) have
interpreted the disk as being aligned parallel to the large scale
outflow. In this case either the source is unrelated to the large
scale outflow, or the orientation has changed in the last $\sim
10^3$~yr, the timescale of current outflow activity. Normally one
would regard this last possibility as extremely unlikely, however, the
motion of a $\sim 10\:M_\odot$ young star (the BN object) through the
region occurred only 500~years ago (Plambeck et al. 1995; Tan 2004b).
Several pieces of evidence point to an ejection of BN from the
already-formed $\Theta^1C$ binary system, however it is not possible
to exclude an origin at source {\it I} itself (Bally \& Zinnecker
2005; Rodriguez et al. 2005). 

Outflows are common from regions of high-mass star formation (see
Beuther \& Shepherd 2005, these proceedings) for a review. However,
because massive stars tend to be forming in clusters it is not always
clear which sources are responsible for driving the outflows.
Nevertheless there seems to be a multitude of collimated, powerful
outflows, that appear to be scaled-up versions of those from low-mass
protostars. The continuity in outflow properties from the low to high
mass regimes suggests that there is a single driving mechanism
(Beuther et al. 2002).

One difference between outflows from low-mass and high-mass protostars
is the presence of high flux of ionizing radiation in the latter. This
should create an ``outflow-confined'', hyper-compact HII region (Tan
\& McKee 2003). This model can account for the radio spectrum and
morphology of source {\it I} in Orion KL, and perhaps also for the
radio sources in CRL~2136 (Menten \& van der Tak 2004), W33A,
AFGL~2591 and NGC~7538~IRS9 (van der Tak \& Menten 2005). An
alternative model is the gravitationally-confined ionized accretion
flow (Keto 2003), however this requires spherical accretion all the
way to the star. Another model is the ionized flow from a
photo-evaporated neutral disk (Hollenbach et al. 1994), however, if
normal MHD outflows are present from the inner disk, they should block
ionization of the outer disk.

\section{The timescale of star cluster formation}

The timescales of star cluster formation have been reviewed by Tan
(2005). Two independent pieces of evidence suggest that in the Orion
Nebula Cluster, stars have been forming for at least 10 dynamical
timescales, or 20 free-fall timescales. First ages of stars derived
from pre-main-sequence tracks show a spread from 0 to at least 3 Myr
(Palla \& Stahler 1999).  Second, the age of a dynamical ejection
event of 4 massive stars ejected from a region coincident with the ONC
is about 2.5~Myr (Hoogerwerf et al. 2001).

A relatively long formation timescale is also consistent with the
observed morphologies of protoclusters in CS molecular lines: Shirley
et al. (2003) find approximately spherical and centrally concentrated
morphologies for a large fraction of their sources, suggesting they
are older than a few dynamical times.

Formation timescales longer than a dynamical time allow the clump gas
to virialize and come into pressure equilibrium: self-gravity is
countered by internal sources of pressure. Numerical simulations (Mac
Low et al. 1998; Stone et al. 1998) suggest that turbulence decays in
one or two dynamical timescales (however, see Cho \& Lazarian 2003).
In this case, in order for turbulence support of the clump to be
maintained, energy must be injected, most probably from internal
sources such as protostellar outflows.

Such long formation timescales would also allow for significant
dynamical relaxation of the forming star cluster: for $N$ equal mass
stars the relaxation time is $t_{\rm relax}\simeq 0.1 N/({\rm ln} N)
t_{\rm dyn}$, i.e. about 14 crossing timescales for $N=1000$.  Using
numerical experiments, Bonnell \& Davies (1998) found that the mass
segregation time (of clusters with mass-independent initial velocity
dispersions) was similar to the relaxation time. The presence of gas
should shorten these timescales (Ostriker 1999). Therefore at least a
part of the observed central concentration of massive stars in the
Orion Nebula Cluster, in particular the Trapezium stars, may be due to
mass segregation rather than preferential formation at the center.

It should be noted that a star cluster formation timescale of a few
Myr is similar to the dynamical timescale of individual GMCs. Star
formation appears to be rapid when compared to these timescales, but
not when compared to the timescales of the star-forming clumps
themselves. This is a major difference between the clustered (e.g.
Orion) and distributed (e.g. Taurus, Hartman 2002) mode of star
formation. Note also that even if the star cluster formation timescale
is similar to the GMC dynamical timescale, this does not imply GMC
lifetimes are this short (e.g. Tassis \& Mouschovias 2004; \S6).

\section{How does feedback affect the formation process?}

Feedback processes that act against gravitational collapse and
accretion of gas to protostars include radiation pressure (transmitted
primarily via dust grains), thermal pressure of ionized regions and
ram pressure from stellar winds, particularly MHD-driven outflows from
protostars that are still actively accreting. If star cluster
formation takes longer then $\sim 3$~Myr, then there is a chance of
supernova feedback clearing out any remaining gas.  

\subsection{Feedback in Individual Cores}

For individual low-mass star formation from a core, bipolar
protostellar outflows, accelerated from the inner accretion disk and
star by rotating magnetic fields, appear to be the dominant feedback
mechanism, probably preventing accretion from polar directions and
also diverting a fraction, up to a third, of the material accreting
through the disk. This leads to star formation efficiencies from the
core of order 50\% (Matzner \& McKee 2000). 

For massive protostars, forming in the same way from a core and
accretion disk, one expects similar MHD-driven outflows to be present
leading to similar formation efficiencies. In addition,
once the massive protostar has contracted to the main sequence (this
can occur rapidly before accretion has finished), it starts to produce
large a flux of ionizing photons. The HII region is unlikely to be
impeded by an accretion flow with reasonable angular momentum.
However, it is likely to be confined, at least equatorially, by the
bipolar outflow (Tan \& McKee 2003). As the protostellar mass and
ionizing flux increase, then eventually the HII region can
spread through the outflow and start to ionize the disk surface. If
the disk is ionized out to a radius where the escape speed is about
equal to the ionized gas sound speed, then a photo-evaporated flow is
set up, further reducing accretion to the star (Hollenbach et
al. 1994).

Radiation pressure on dust grains (well-coupled to the gas
at these densities) is also important for massive protostars. It has
been suggested, in the context of spherical accretion models,
that this leads to an upper limit to the initial mass function (Kahn
1974; Wolfire \& Cassinelli 1987). However, these constraints are
relaxed once a disk geometry is allowed for (Nakano 1989; Jijina \&
Adams 1996). Yorke \& Sonnhalter (2002) used 2D axially symmetric
simulations to follow massive star formation from a core collapsing to
a disk, including radiation pressure feedback: accretion stopped at
$43\:M_\odot$ in their most massive core.  They showed the accretion
geometry channeled radiative flux into the polar directions and away
from the disk, terming this the ``flashlight effect''. Krumholz,
McKee, \& Klein (2005b) found that cavities created by protostellar
outflows increase the flashlight effect, allowing even higher final
masses.

\subsection{Feedback during Star Cluster Formation}

One of the primary goals for models of feedback in star clusters is a
prediction of the star formation efficiency, since this determines
whether the cluster remains bound: Lada, Margulis, \& Dearborn (1984)
find from numerical models that clusters can remain bound with
efficiencies as low as $\sim 30\%$ if the gas is removed gradually.
Lada \& Lada (2003) conclude that 90-95\% of Galactic embedded
clusters emerge from their GMCs unbound, although the mass associated
with these systems is a somewhat lower percentage.

One can make some simple analytic estimates of the effects of massive
star feedback on real protoclusters using Figure~1. The dashed line
shows the condition that the escape speed at a
distance $2R$ from the clump center is equal to the ionized gas sound
speed ($\sim 10\:{\rm km\:s^{-1}}$). To the right and above this line
ionizing feedback is much less effective since even if cluster gas is
ionized it will be relatively difficult to be expelled. The Arches
cluster in the Galactic center and super star clusters are in this
region. If star clusters form relatively slowly, e.g. in 20$t_{\rm
  ff}$ as may be the case in Orion (\S5), then we can see that the
clusters forming massive stars, which have approximately constant
densities and free-fall timescales ($\sim 10^5\:{\rm yr}$), would not
be affected by supernova feedback, since this only starts after at
least $\sim 3\times 10^6\:{\rm yr}$. However, the rate of star
formation is uncertain, particularly in the more massive and distant
clusters.

Matzner \& McKee (2000) modeled protostellar outflow feedback in
clusters of low-mass stars, estimating formation efficiencies of
30-50\%. Adding high-mass stars to these particular models would
presumably reduce the efficiency.

Tenorio-Tagle et al. (2003) presented a 1D model of star cluster
formation in the presence of stellar wind and ionizing feedback,
assuming an initial burst of massive star formation that creates a
compressed shell from the infalling neutral gas, where more stars can
form. They achieved high star formation efficiencies, allowing the
build-up of very massive clusters, comparable to super star clusters.
However, it is not clear how their model would fare in a more
realistic turbulent and clumpy medium.

Scoville et al. (2001) considered radiation pressure feedback as a
mechanism for limiting star cluster masses at $\sim
10^3\:M_\odot$. However, their model is 1D and it would be more
difficult to disrupt gas if it were in optically thick clumps.

Tan \& McKee (2001; 2004) used an idealized model to investigate
feedback in a turbulent and clumpy medium.  This structure was
approximated by dividing the gas into cores and an intercore medium.
The dynamics of the cores are affected by the potential of the overall
protocluster and feedback effects from a stellar population at the
cluster center, including radiation pressure, stellar winds and
ionization, which can photo-evaporate cores. The main conclusion was
that a clumpy, turbulent medium is much more capable of confining
feedback, particularly ionizing feedback. HII regions are confined
because they are continually injected with neutral cores that then
suffer high photo-evaporation rates. This mass loading keeps the
density and recombination rate relatively high. Such effects are
likely to be important for Galactic ultra-compact ($\lesssim 0.1$~pc)
HII regions, whose long lifetimes have been a puzzle. Figure~1 shows
the parameters of three models, A, B, and C. Model A formed a cluster
that dispersed its gas in 2~Myr, while B and C took about 3~Myr.
Estimates of the star formation efficiency are somewhat uncertain as
protostellar outflows were omitted. Within the limitations of the
model, the efficiencies were $\sim 30\%$ for model A, and $\sim 50\%$
for models B and C.

\subsection{Feedback on GMCs}

Once star clusters have formed, their feedback will impact the larger
scale interstellar medium. In particular they may contribute to the
destruction of GMCs. Williams \& McKee (1997) considered the
destruction of Galactic GMCs by photo-evaporation from ionizing
photons from OB associations, finding destruction timescales of $\sim
30$~Myr for the most massive clouds. Matzner (2002) estimated slightly
shorter destruction timescales for the same feedback process. Monaco
(2004) considered the effects of supernova feedback on clouds that
have already been shaped by ionization. 

Alternatively, Ballesteros-Paredes (2004) argued that GMCs are
transient phenomena and that their disruption is simply due to
dynamical processes. Clark \& Bonnell (2004) modeled star
formation in such transient GMCs.

Observationally, Leisawitz, Bash \& Thaddeus (1989) found that open
star clusters older than about $\sim 10$~Myr were not associated with
molecular clouds, which is consistent either with post-star-formation
cloud lifetimes shorter than this age, i.e. only a couple of dynamical
timescales, or with relative velocities of star clusters and their
parent clouds of about $10\:{\rm km\:s^{-1}}$, which are to be
expected from photoionization feedback (Williams \& McKee 1997).
The important question of GMC lifetimes remains open.

\begin{acknowledgments}
  I thank many colleagues for discussions, particularly Chris McKee
  and Mark Krumholz.  I am supported by a Zwicky fellowship from ETH.
\end{acknowledgments}

\bibliographystyle{kapalike}







%





\end{document}